\documentclass[prl, aps, twocolumn, floatfix, showpacs]{revtex4}
\usepackage{graphicx, amsmath, amssymb, times}

\begin{document}

\title{Opposite effect of spin-orbit coupling on condensation and superfluidity}
\author{Kezhao Zhou}
\affiliation{Shenyang National Laboratory for Materials Science,
Institute of Metal Research, and International Centre for Materials
Physics, Chinese Academy of Sciences, 72 Wenhua Road, Shenyang
110016, People's Republic of China}
\author{Zhidong Zhang}
\affiliation{Shenyang National Laboratory for Materials Science,
Institute of Metal Research, and International Centre for Materials
Physics, Chinese Academy of Sciences, 72 Wenhua Road, Shenyang
110016, People's Republic of China}
\date{\today}

\begin{abstract}
We investigated effects of a Rashba-type spin-orbit coupling (SOC)
on the condensed density and superfluid density tensor of a
two-component Fermi gas in the BCS-BEC crossover at zero
temperature. In anisotropic three dimensions (3D), we found that SOC
has an opposite effect on condensation (enhanced) and superfluidity
(suppressed in the SOC direction) and this effect becomes most
pronounced for very weak interactions and the SOC strength being
larger than a characteristic value. Furthermore, as functions of SOC
strength, the condensed density changes monotonously for all
interaction parameters while the superfluid density has a minimum
when the interaction parameter is below a critical value. We also
discussed the isotropic two dimensional (2D) case where analytical
expressions for the gap and number equations were obtained and the
same phenomena was found as that of the 3D case.
\end{abstract}

 \maketitle

\textit{Introduction.} 
Spin-orbit coupling (SOC) is a central topic in condensed matter
physics. First, it plays an essential role for the realization of
nontrivial topological states which are discussed intensively
nowadays \cite{kane}. Secondly, as was shown by Gor'kov and Rashba
\cite{rashba}, SOC can induce a nontrivial spin-triplet pairing
field which leads to significant changes in the properties of
superconductors \cite{bergeret}. Quite recently, effective SOC was
realized for bosonic $^{87}Rb$ ultra-cold atoms by dressing two
atomic spin states with a pair of lasers \cite{nist}. With the
anticipation that this novel technique is also applicable to Fermi
atoms, practical proposal of generating SOC in fermionic $^{40}K$
atoms with tunable interaction through Feshbach Resonance is given
in \cite{sau}.

Motivated by this new progress, effect of SOC on the pairing and
superfluid nature of Fermi systems in the BCS-BEC crossover has
become a cutting-edge field recently because of its broad interests
in condensed matter physics. The spin-triplet pairing fields and
anisotropic nature of the superfluidity induced by SOC were
investigated in \cite{vs1} and proposal for detecting this
phenomenon was given in \cite {hu} through measurement of the
momentum distribution and single-particle spectral function. On the
other hand, SOC significantly enhances the pairing phenomena as was
shown by the exact two-body solutions \cite{vs2} where a new bound
state (rashbons) emerges and many-body mean-field calculations
\cite{han,iskin,zhai}.

In this Letter, we study the effects of SOC on two fundamental
quantities: condensation and superfluidity. Condensation is well
described by the concept of off-diagonal-long-range-order
\cite{yang}. However, Landau's approach of calculation of superfluid
density (tensor) is only applicable to systems satisfying Galilean
transformation \cite{landau}. For systems in the presence of SOC
obviously violating Galilean transformation, we gave the general
method of calculating the superfluid density tensor. Furthermore, we
found that at zero temperature, SOC enhances condensation while
suppresses superfluidity in both 3D and 2D. Up to our knowledge,
this is the first demonstration of such opposite behaviors of
condensation and superfluidity driven by SOC and renews our previous
knowledge that these two phenomena change in the same direction with
other influencing factors (such as temperature and disorder).

\textit{The model.}
In the presence of SOC, the system of a two-component Fermi gas can
be
described by the finite temperature grand-partition function $Z=\int d[\bar{%
\psi}_{\sigma },\psi _{\sigma }]\exp \left( -S[\bar{\psi}_{\sigma
},\psi
_{\sigma }]\right) $ ($\hbar =k_{B}=1$ through out) where the action $S[\bar{%
\psi}_{\sigma },\psi _{\sigma }]$ is given by
$S[\bar{\psi}_{\sigma},\psi
_{\sigma }]=\int_{0}^{\beta }d\tau \int d^{d}\mathbf{r}\sum_{\sigma }[\bar{%
\psi}_{\sigma }\partial _{\tau }\psi _{\sigma }+\mathcal{H}_{0}+\mathcal{H}%
_{I}]$ with $\beta =1/T$ , $\sigma =\uparrow ,\downarrow $ denoting spin, $%
\bar{\psi}_{\sigma },\psi _{\sigma }$ being the Grassmann fields,
and $d\left(
=2,3\right) $ being the dimension. We focus on a Rashba-type SOC \cite%
{rashba} and the single particle Hamiltonian density can be written as $%
\mathcal{H(}\bar{\psi},\psi )\mathcal{=}\bar{\psi}\left( \hat{\xi}_{\mathbf{p%
}}+\mathcal{H}_{so}\right) \psi $ where $\psi =[\psi _{\uparrow
},\psi _{\downarrow }]^{T}$ is the collective fermionic field, the
kinetic operator
$\hat{\xi}_{\mathbf{p}}=\mathbf{\hat{p}}^{2}/(2m)-\mu $ with $\mu $
being the chemical potential, the Rashba term
$\mathcal{H}_{so}=\lambda \left( \mathbf{\hat{\sigma}\times
\hat{p}}\right) _{z}$ with $\hat{\sigma}$ being the Pauli matrices
and $\lambda $ being the SOC strength. The singlet-channel
attractive interaction can be characterized by a contact
interaction parameter $g\left( <0\right) $ and correspondingly $\mathcal{H}%
_{I}=g\bar{\psi}_{\uparrow }\bar{\psi}_{\downarrow }\psi
_{\downarrow }\psi _{\uparrow }$.

In order to study the Fermi-pairing problems, we employ the
Hubbard-Stratonovich transformation \cite{altland} to cancel the
four-body interaction term $\mathcal{H}_{I}$ by introducing a
pairing field $\Delta \left( \mathbf{r},\tau \right) $. After
integrating out the fermionic
fields, we obtain the effective pairing action as $S_{eff}\left[ \bar{\Delta}%
,\Delta \right] =-\int_{0}^{\beta }d\tau \int
d^{d}\mathbf{r}\left\vert \Delta \left( \mathbf{r},\tau \right)
\right\vert ^{2}/g-1/2Tr\ln \left[
\mathcal{G}_{\mathbf{r},\tau }^{-1}\right] $ where the inverse propagator $%
\mathcal{G}_{\mathbf{r},\tau }^{-1}$ is
\begin{equation}
\mathcal{G}_{\mathbf{r},\tau }^{-1}=\left[
\begin{array}{cccc}
\partial _{\tau }+\hat{\xi}_{\mathbf{p}} & \hat{\gamma}_{_{\mathbf{p}}} & 0
& \Delta \\
\hat{\gamma}_{_{\mathbf{p}}}^{\ast } & \partial _{\tau }+\hat{\xi}_{\mathbf{p%
}} & -\Delta & 0 \\
0 & -\bar{\Delta} & \partial _{\tau }-\hat{\xi}_{\mathbf{p}} & \hat{\gamma}%
_{_{\mathbf{p}}}^{\ast } \\
\bar{\Delta} & 0 & \hat{\gamma}_{_{\mathbf{p}}} & \partial _{\tau }-\hat{\xi}%
_{\mathbf{p}}%
\end{array}%
\right] \text{,}  \label{inversepropagator}
\end{equation}%
with $\hat{\gamma}_{_{\mathbf{p}}}=\lambda \left( \hat{p}_{y}+i\hat{p}%
_{x}\right) $.

\textit{Mean-field theory}.
At the mean-field level $\Delta \left( \mathbf{r},\tau \right)
=\Delta _{0}$ which is referred to as the gap parameter, the
effective pairing action becomes $S_{eff}\left[ \bar{\Delta},\Delta
\right] =-\beta V\Delta
_{0}^{2}/g-1/2\sum_{\mathbf{p},i\omega _{n}}\ln \left[ \det \mathcal{G}_{%
\mathbf{p},i\omega _{n}}^{-1}\right] $ where $\mathcal{G}_{\mathbf{p}%
,i\omega _{n}}^{-1}$ is the momentum-frequency representation of Eq. ({\ref%
{inversepropagator}}) with $\Delta \left( \mathbf{r},\tau \right)
=\Delta _{0}$, $V$ is the size of the system and $\omega _{n}=\left(
2n+1\right) \pi
/\beta $ are the Fermi Matsubara frequencies. From $\det \mathcal{G}_{%
\mathbf{p},E}^{-1}=0,$ the excitation spectrum can be obtained as $E_{%
\mathbf{p,}\delta }=\sqrt{\left( \xi _{\mathbf{p}}+\delta \left\vert
\gamma
_{\mathbf{p}}\right\vert \right) ^{2}+\Delta _{0}^{2}}$ and $E_{\mathbf{p,}%
\delta }^{\prime }=-E_{\mathbf{p,}\delta }$ where $\xi _{\mathbf{p}%
}=\epsilon _{\mathbf{p}}-\mu $ with $\epsilon _{\mathbf{p}}=\mathbf{p}%
^{2}/2m $ and $\delta =\pm 1$ is called helicity. Finally, by using
the thermodynamic relation $\Omega =-1/\beta \ln Z_{0}$, we have the
thermodynamic potential $\Omega _{0}=-V\Delta _{0}^{2}/g+1/2\sum_{\mathbf{%
\mathbf{p},\delta }}\left( \xi _{\mathbf{p}}-E_{\mathbf{p,}\delta
}\right)
-1/\beta \sum_{\mathbf{p},\delta =\pm }\ln \left( 1+e^{-\beta E_{\mathbf{p}%
,\delta }}\right) $ from which the gap and number equations are
given by
\begin{equation}
\frac{1}{g}=-\frac{1}{V}\sum_{\mathbf{p,}\delta =\pm }\frac{\tanh
\left( \frac{\beta E_{\mathbf{p,}\delta }}{2}\right)
}{4E_{\mathbf{p,}\delta }}, \label{gap}
\end{equation}%
\begin{equation}
n=\frac{1}{2V}\sum_{\mathbf{p,}\delta =\pm }\left[ 1-\frac{\left( \xi _{%
\mathbf{p}}+\delta \left\vert \gamma _{\mathbf{p}}\right\vert
\right) \tanh
\left( \frac{\beta E_{\mathbf{p,}\delta }}{2}\right) }{E_{\mathbf{p,}\delta }%
}\right] .  \label{num}
\end{equation}

Eq. ({\ref{gap}}) and Eq. ({\ref{num}}) are the generalized BCS gap
and number equations in the presence of a Rashba-type SOC which have
been investigated in detail to study the ground state and finite
temperature properties of this novel system. The key discovery is
that the increased density of states by SOC plays a crucial role for
the understanding of the pairing enhancing phenomena \cite{zhai}.
With these results in mind, we now move on to the calculation and
discussion of condensed density and superfluid density tensor.

\textit{Condensed density}.
For the Fermi pairing problems, the condensed density is generally
defined as \cite{yang} $n_{c}=1/V\sum_{\mathbf{p,}ss^{\prime
}}\left\vert \left\langle
\bar{\psi}_{\mathbf{p},s}\psi_{-\mathbf{p},s^{\prime }}\right\rangle
\right\vert ^{2}$. For the system considered in this paper, the
singlet-channel attractive interaction supports a singlet-pairing
field while SOC can simultaneously induce a triplet component.
Within the mean-field theory, spin-singlet and -triplet pairing
fields are given by
\cite{hu}: $\left\langle \bar{\psi}_{\mathbf{p,}\uparrow }\psi_{-%
\mathbf{p,}\downarrow }\right\rangle =\Delta _{0}\sum_{\delta }\tanh
\left( \beta E_{\mathbf{p},\delta }/2\right) /\left(
4E_{\mathbf{p},\delta }\right)
$ and $\left\langle \bar{\psi}_{\mathbf{p,}\uparrow }\psi_{-\mathbf{p,}%
\uparrow }\right\rangle =-\Delta _{0}\left( \gamma
_{\mathbf{p}}/\left\vert \gamma _{\mathbf{p}}\right\vert \right)
\sum_{\delta }\delta \tanh \left( \beta E_{\mathbf{p},\delta
}/2\right) /\left( 4E_{\mathbf{p},\delta }\right) $, respectively.
The spin-singlet contribution to the condensed fraction was first
discussed in \cite{sala1} where it was shown to behave
non-monotonously with a minimum as a function of SOC strength for
weak enough interaction parameter. In this Letter, we take both
pairing components into consideration and the full condensed density
becomes
\begin{equation}
n_{c}=\frac{\Delta _{0}^{2}}{4}\frac{1}{V}\sum_{\mathbf{p,\delta }}\frac{%
\tanh ^{2}\left( \frac{\beta E_{\mathbf{p},\delta }}{2}\right) }{E_{\mathbf{p%
},\delta }^{2}}.  \label{condensed}
\end{equation}

At zero temperature, repulsive interactions between Fermi pairs
(Bosons) result in depletion of the condensate which is a familiar
phenomenon for interacting BEC systems.

\textit{Superfluid density}.
Unlike the condensate density, the superfluidity is a kinetic
property of the system. By Landau's theory \cite{landau}, the normal
mass of the system can be obtained through the calculation of the
total momentum carried by excitations when the system is enforced in
a uniform superfluid flow with velocity $\mathbf{v}_{s}$
\begin{equation}
\mathbf{P}=\sum_{\mathbf{p,}\sigma }\mathbf{p}f\left( E_{\mathbf{p,}\sigma }-%
\mathbf{p\cdot v}_{s}\right) ,  \label{landau}
\end{equation}%
where $\sigma $ is a conserved quantum number which is spin in the
absence of SOC, $f\left( x\right) =1/\left( e^{\beta x}\pm 1\right)
$ is the Fermi/Bose distribution function depending on the nature of
the excitations, and $E_{\mathbf{p,}\sigma }-\mathbf{p\cdot v}_{s}$
is the excitation spectrum for moving system obtained from Galilean
transformation. At zero temperature, no excitations are created at
very small $\mathbf{v}_{s}$ and the superfluid density coincides
with the total density.

However the situation is dramatically changed in the presence of SOC
where Galilean transformation is violated. In order to calculate the
response of the system to a uniform superfluid flow in the presence
of SOC, instead of using Eq. ({\ref{landau}}) which is no
longer valid, we calculate the increasing of thermodynamic potential $%
\delta \Omega =\Omega \left( \mathbf{v}_{s}\right) -\Omega \left(
0\right) =\left( 1/2\right) V\sum_{\alpha \eta }mn_{s,\alpha \eta
}v_{\alpha }v_{\eta }$ from which $n_{s,\alpha \eta }=1/\left(
mV\right) \left[ \partial ^{2}\Omega \left( \mathbf{v}_{s}\right)
/\partial v_{s,\alpha }\partial v_{s,\eta }\right]
_{\mathbf{v}_{s}=0}$ is defined as the superfluid density tensor
\cite{taylor}. A convenient approach of generating such superfluid
flow is applying a ``phase twist" to the order parameter \cite{fisher}: $%
\Delta \left( \mathbf{r},\tau \right) =\Delta _{0}e^{i\mathbf{q\cdot
r}}$ and correspondingly $\mathbf{v}_{s}=\mathbf{q}/2m$. Therefore
the superfluid density tensor can be defined as $n_{s,\alpha \eta
}=4m\left[ \partial
^{2}\Omega \left( \mathbf{q}\right) /\partial q_{\alpha }\partial q_{\eta }%
\right] _{\mathbf{q}=0}$.

By substitution of $\Delta \left( \mathbf{r},\tau \right) =\Delta _{0}e^{i%
\mathbf{q\cdot r}}$ into Eq. ({\ref{inversepropagator}}), the
thermodynamic
potential for moving system can be obtained as $\Omega \left( \mathbf{q}%
\right) =-V\Delta _{0}^{2}/g+\sum_{\mathbf{k}}\left( \tilde{\xi}_{\mathbf{k}%
}-\mathbf{k\cdot q/}2m\right) -1/\left( 2\beta \right) \sum_{\mathbf{p,}%
i=1\sim 4}\ln \left[ 1+e^{\beta \left( \tilde{E}_{\mathbf{p,}i}-\mathbf{%
k\cdot q/}2m\right) }\right] $ where $\tilde{\xi}_{\mathbf{p}}=\xi _{\mathbf{%
p}}+\mathbf{q}^{2}/\left( 8m\right) $ and $\tilde{E}_{\mathbf{p},i}$
are solutions of
\begin{equation}
\left( \tilde{E}_{\mathbf{p},i}^{2}-\omega _{\mathbf{p}}\right)
^{2}-4\left\vert \tilde{E}_{\mathbf{p},i}\gamma _{\mathbf{q/2}}-\tilde{\xi}_{%
\mathbf{p}}\gamma _{\mathbf{p}}\right\vert ^{2}+\Lambda
_{\mathbf{p}}^{2}=0, \label{eigen}
\end{equation}%
with $\omega _{\mathbf{p}}=\tilde{\xi}_{\mathbf{p}}^{2}+\Delta
_{0}^{2}+\left\vert \gamma _{\mathbf{p}}\right\vert ^{2}-\left\vert \gamma _{%
\mathbf{q/2}}\right\vert ^{2}$, and $\Lambda _{\mathbf{p}}=Im\left( \gamma _{%
\mathbf{p}}\gamma _{\mathbf{q}}^{\ast }\right) $. In the presence of
SOC, when the whole system is moving with a uniform velocity, the
original four excitation spectrums ($E_{\mathbf{p,}\delta
},-E_{\mathbf{p,}\delta }$) are strongly coupled and correspondingly
Eq. ({\ref{landau}}) is not well defined now. Superfluidity in
systems that violates the Galilean transformation has also been
discussed in the bosonic systems in the presence of SOC where the
critical velocity has been discussed with the same method used in
the calculation of the excitation spectrum for moving systems
\cite{wu}.

Combined with Eq. ({\ref{eigen}),} calculation of the second-order
derivative of $\Omega \left( \mathbf{q}\right) $ with respect to
$q_{i}$ is straightforward although tedious. Finally, the superfluid
density tensor is obtained as
\begin{equation}
n_{s,zz}=\frac{N}{V}-\frac{4m}{V}\sum_{\mathbf{p},\delta }Y_{\mathbf{p,}%
\delta }\left( \frac{p_{z}}{2m}\right) ^{2},  \label{nsz}
\end{equation}%
\begin{eqnarray}
n_{s,\parallel } &=&\frac{N}{V}-\frac{2m}{V}\sum_{\mathbf{k},\delta }Y_{%
\mathbf{p,}\delta }\left( \frac{\left\vert \gamma _{\mathbf{p}}\right\vert }{%
2m\lambda }+\delta \frac{\lambda }{2}\right) ^{2}  \notag \\
&&-\frac{m\lambda ^{2}}{4V}\sum_{\mathbf{p},\delta }\tanh \left(
\frac{\beta
E_{\mathbf{p,}\delta }}{2}\right) \frac{\xi _{\mathbf{p}}^{2}+\delta \xi _{%
\mathbf{p}}\left\vert \gamma _{\mathbf{p}}\right\vert +\Delta _{0}^{2}}{%
\delta \xi _{\mathbf{p}}\left\vert \gamma _{\mathbf{p}}\right\vert E_{%
\mathbf{p,}\delta }},  \label{nsxy}
\end{eqnarray}%
where $Y_{\mathbf{p},\delta }=\beta f\left( E_{\mathbf{p,}\delta }\right) %
\left[ 1-f\left( E_{\mathbf{p,}\delta }\right) \right] \ $with
$f\left(
x\right) $ being the Fermi distribution function, $n_{s,xx}=n_{s,yy}=n_{s,%
\parallel }$, and $n_{s,\alpha \neq \eta }=0$. In 3D, the anisotropic nature
of the superfluid can be evidently seen from $n_{s,zz}\neq n_{s,\parallel }$%
. Since SOC does not affect motion in $z$ direction, spin is a
conserved
quantum number and $n_{s,zz}$ has the same form as that obtained from Eq. ({%
\ref{landau}}). However, superfluid motion in $x,y$ direction is
dramatically changed by SOC. Most interestingly, a new term (last
line in Eq. ({\ref{nsxy}})) emerges due to SOC and it is not zero at
$T=0$ which means suppression of superfluidity in $x,y$ direction.
The first line of Eq. ({\ref{nsxy}}) can be understood in spirit of
Landau's theory where momentum carried by excitations
$E_{\mathbf{p,}\delta }$ is now shifted by $\delta \lambda m$ due to
SOC. At $T=T_{c}$ where $\Delta _{0}=0,$ both $n_{s,z}$ and
$n_{s,\parallel }$ equal to zero which is crucial for the
correctness of our results. In 2D, the system is isotropic where the
superfluid density is only given by $n_{s,\parallel }$.

With the anticipation that the mean-field theory is sufficient to
capture the qualitatively correct physics in the whole BCS-BEC
region at zero temperature as demonstrated in \cite{giorgini} in the
absence of SOC, we only focus on the zero temperature behaviors of
the condensed and superfluid density.

\textit{Results and discussion.}
Because, at zero temperature, behaviors of the condensed and
superfluid density ($n_{s,\parallel}$) are the same in 3D and 2D
cases with only quantitative differences, we will discuss the 3D
case in detail and give a brief discussion on the 2D results.

As usual we regularize the contact interaction parameter $g$ in Eq. ({\ref%
{gap}}) by the experimentally related scattering length $a$ through $%
1/g=m/\left( 4\pi a\right) -1/V\sum_{\mathbf{p}}1/\left( 2\epsilon _{\mathbf{%
p}}\right) $. With the gap and chemical potential obtained from the
self-consistent solutions of Eq. ({\ref{gap}}) and Eq.
({\ref{num}}), we numerically calculate Eq. ({\ref{condensed}}) and
Eq. ({\ref{nsxy}}) and the
results are shown in Fig.1. %
\begin{figure}[tbh]
\includegraphics[width=\columnwidth,height=80mm]{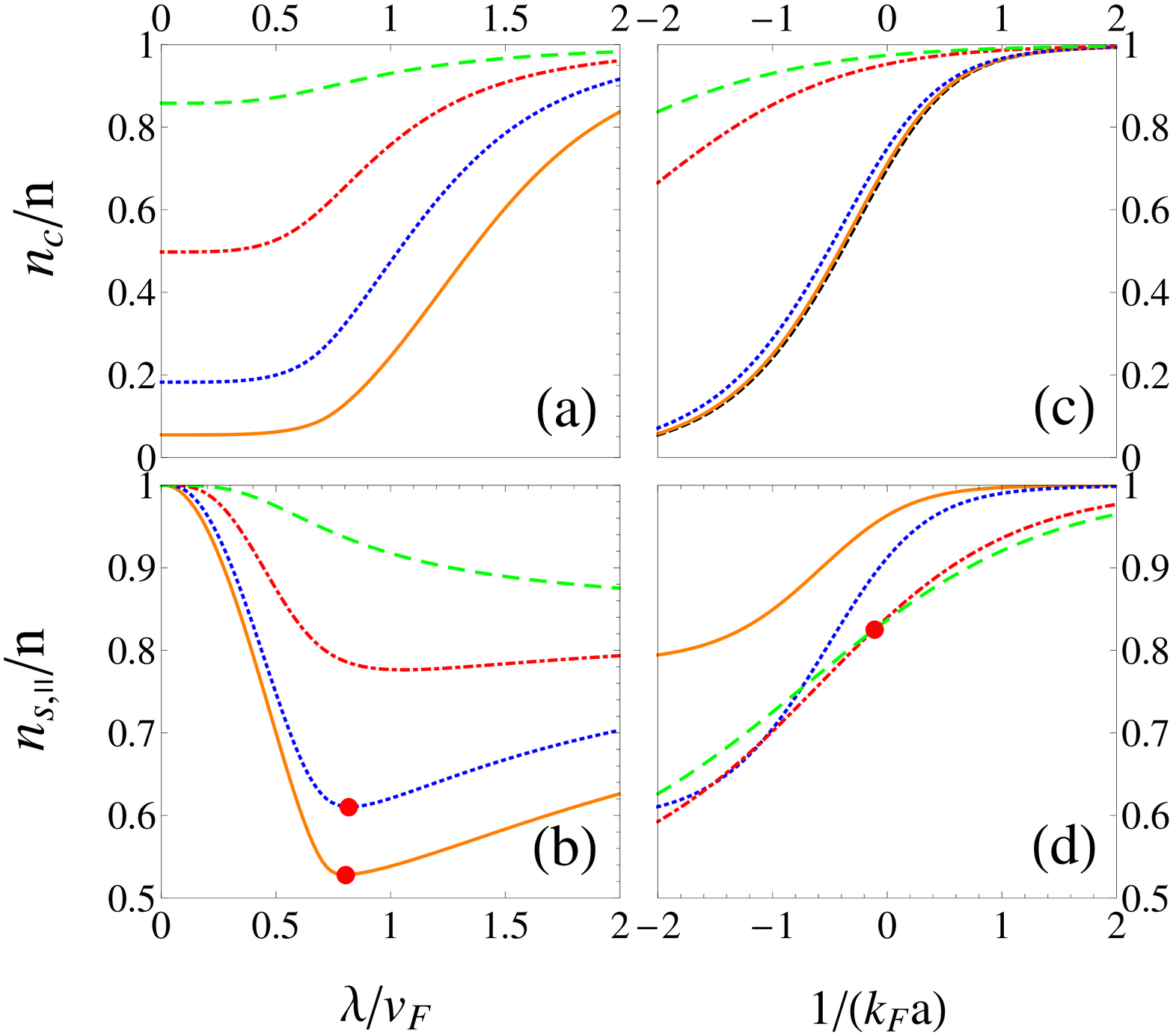}
\caption{(Color online) Condensed and superfluid fraction as functions of $%
\protect\lambda /v_{F}$ and $1/(k_{F}a)$ in 3D. In (a) and (b),
$1/(k_{F}a)$ are given as: $-2$, $-1.2$, $-0.4$ and $0.4$ for the
same line species from below. In (c) and (d), $\protect\lambda
/v_{F}$ are set to be 0.4 for solid orange lines, 0.6 for dotted
blue lines, 1.6 for dot-dashed red lines and 2.0
for media-dashed green lines. Dashed black line in (c) corresponds to $%
\protect\lambda =0$. Red points in (b) and (d) are explained in
text.} \label{3dnsnc}
\end{figure}
As can be seen from Fig. \ref{3dnsnc}(a) and (c), the condensed
density is always enhanced by SOC. Nevertheless, seen from Fig.
\ref{3dnsnc}(a), we can still define a characteristic value roughly
located at $\lambda _{c}\approx
0.5v_{F}$ where $v_{F}=k_{F}/m$ and $k_{F}$ is defined through $%
k_{F}^{2}=3\pi ^{2}n$. For $\lambda <\lambda _{c}$, condensed
fraction defined as $n_{c}/n$ increases only slightly which can also
be seen from Fig. \ref{3dnsnc}(c) that the solid orange and dotted
blue lines almost coincide with
the dash black line (where $\lambda =0$). Only when $\lambda >\lambda _{c}$ can $%
n_{c}/n$ have a significant increase. For $1/k_{F}a\rightarrow
+\infty $, the effect of SOC becomes very weak and
$n_{c}/n\rightarrow 1$. For $\lambda _{c}\gg v_{F}$, we also have
$n_{c}/n\rightarrow 1$ which agrees with the statement that SOC can
produce a bound state and thus induces a crossover even for very
weak interactions \cite{vs2, zhai}.

On the contrary, the superfluidity is always suppressed by SOC as
can be seen in Fig. \ref{3dnsnc}(b). Furthermore, as a function of
$\lambda $, superfluid fraction $n_{s,\parallel }/n$ varies
non-monotonously with a
minimum located at $\lambda _{c}^{\prime }$ (denoted by red points in Fig. %
\ref{3dnsnc}(b)). This minimum exists for interaction parameter
below a well-defined critical value $1/k_{F}a_{c}\approx -0.13$
which can be determined in Fig. \ref{3dnsnc}(d) as the rightmost red
crossing point. However, we find that SOC never destroys the
superfluid completely for all interaction parameters. The minimum
superfluid fraction $\left( n_{s,\parallel }/n\right) _{\min
}\rightarrow 0.5$ for $1/k_{F}a\rightarrow -\infty $. When
$1/k_{F}a>1/k_{F}a_{c}$, this monotonic behavior disappears and
$n_{s,\parallel }/n$ decreases with increasing $\lambda $.

The opposite behavior of condensation and superfluidity controlled
by SOC strength is found to be most pronounced for very weak
interaction parameters and $\lambda >\max (\lambda _{c},\lambda
_{c}^{\prime })$. As can be seen from the solid orange lines in Fig.
\ref{3dnsnc}(a) and (b), for $\lambda=2v_{F}$, the condensed
fraction $n_{c}/n$ is increased by $0.78$ and the superfluid density
is suppressed by $0.38 $. For $\lambda <\max (\lambda _{c},\lambda
_{c}^{\prime })$, the superfluid density decreases quickly to its
minimum value while the condensed density changes only very
slightly.

\begin{figure}[tbh]
\includegraphics[width=\columnwidth,height=80mm]{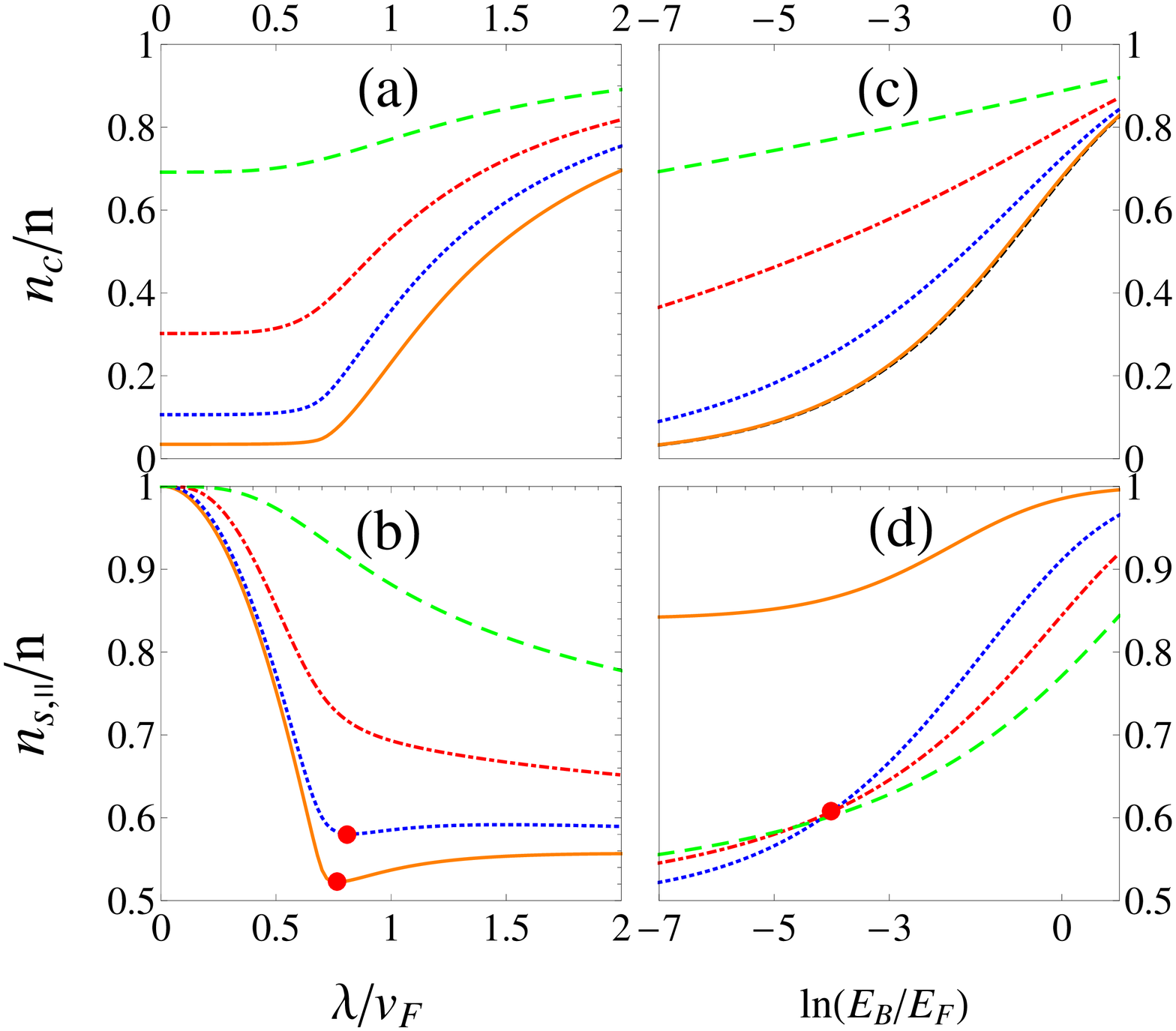}
\caption{(Color online) Condensed and superfluid fraction of 2D
Fermi gas as
functions of $\protect\lambda /v_{F}$ and $E_{B}/E_{F}$. In (a) and (b), $%
E_{B}/E_{F}$ are given as: $0.001$ for solid orange lines, $0.01$
for dotted blue lines, $0.1$ for dot-dashed red lines and $1.1$ for
medium-dashed green lines. In (c) and (d), $\protect\lambda /v_{F}$
are set to be 0.4, 0.8, 1.2 and 2.0 for solid orange, dotted blue,
dot-dashed red, medium-dashed green lines, respectively. Dashed
black line in (c) corresponds to $\protect\lambda =0$. Red points in
(b) and (d) are explained in text.} \label{2dnsnc}
\end{figure}
In 2D, divergence of the integral over momenta can be cured by substituting $%
1/g=-1/V\sum_{\mathbf{p}}1/\left( 2\epsilon
_{\mathbf{p}}+E_{B}\right) $ into Eq. ({\ref{gap}}) where $E_{B}$ is
the binding energy and becomes the controlling parameter for the
BCS-BEC crossover problem. With dimensionless
parameter given by $\tilde{\lambda}=\lambda /v_{F},$ $\tilde{\mu}=\mu /E_{F},%
\tilde{\Delta}_{0}=\Delta _{0}/E_{F}$, $\tilde{E}_{B}=E_{B}/E_{F}$ where $%
E_{F}=k_{F}^{2}/2m$ and $k_{F}$ is defined through $k_{F}^{2}=2\pi n$, Eq. ({%
\ref{gap}}) and Eq. ({\ref{num}}) have analytical expressions as
$1=\left(
\tilde{\mu}+\sqrt{\tilde{\mu}^{2}+\tilde{\Delta}_{0}^{2}}\right) /2+\tilde{%
\lambda}^{2}\left( 1+\tilde{\mu}/\sqrt{\tilde{\mu}^{2}+\tilde{\Delta}_{0}^{2}%
}\right) +H\left( \tilde{\mu},\tilde{\Delta}_{0},\tilde{\lambda}\right) $, $%
\ln E_{b}$ $=\ln \left( \sqrt{\tilde{\mu}^{2}+\tilde{\Delta}_{0}^{2}}-\tilde{%
\mu}\right) -K\left(
\tilde{\mu},\tilde{\Delta}_{0},\tilde{\lambda}\right) $
respectively with $H\left( \tilde{\mu},\tilde{\Delta}_{0},\tilde{\lambda}%
\right) $ and $K\left(
\tilde{\mu},\tilde{\Delta}_{0},\tilde{\lambda}\right) $ given in
\cite{note}. Results of superfluid and condensed fraction are shown
in Fig.2. The same phenomena discussed in 3D are also found in 2D
case. However, we find that $\lambda _{c}$ drifts leftwards when increasing $%
E_{B}/E_{F}$ as can be seen from Fig. \ref{2dnsnc}(a). The critical
interaction parameter for the appearance of a minimum point of $%
n_{s,\parallel }/n$ is $E_{B}\approx 0.018E_{F}$.

\textit{Conclusions.}
In summary, general formulae were obtained for the condensed density
and superfluid density tensor of a two-component Fermi gases in the
presence of a Rashba-type SOC. At zero temperature, we found that
superfluidity in the SOC direction is suppressed while condensation
is enhanced by SOC and this phenomenon becomes most pronounced for
very weak interaction parameters and $\lambda >\max (\lambda
_{c},\lambda _{c}^{\prime })$. Furthermore, the superfluid fraction
exhibits a non-monotonic behavior with a minimum as a function of
SOC strength when interaction parameters is below a critical value
while the condensed fraction increases only monotonously with either
interaction parameter or SOC strength. These phenomena happen in
both 3D and 2D cases.

Finally, we point out that there is an essential difference
considering the mechanism of suppressing superfluidity by disorder
\cite{huang,orso} and SOC. Superfluid motion is suppressed through
energy dissipation due to scattering with impurities. While the
nonzero normal density at zero temperature induced by SOC is a
direct result of the triplet pairing field. In \cite{han}, it is
demonstrated that, at zero temperature, this triplet pairing field
induces a non-zero spin susceptibility which implies a residual
normal fluid \cite{leggett}. An interesting future work is the
combined effect of disorder and SOC on these two phenomena where a
bose-glass state \cite{huang} may show up based on the conclusion
that disorder and SOC both suppress superfluidity while depletion of
the condensate induced by disorder may be weakened/cancelled by SOC.

\textit{Acknowledgements.}
We thank Prof. L. Salasnich for helpful discussions. This work has
been supported by the National Natural Science Foundation of China
under Grant 50831006 and the National Basic Research Program
No.2010CB934603, the Ministry of Science and Technology of China.

\textbf{Note Added:} After finishing this paper, we note that the
full condensed density is also discussed in \cite{sala2}.


\begin{thebibliography}{99}
\bibitem{note} {$\ H\left( \tilde{\mu},\tilde{\Delta},\tilde{\lambda}\right)
=\tilde{\lambda}\tilde{\Delta} ^{2}\left[ N\left( 2\theta ,\kappa
\right) -M\left( 2\theta ,\kappa \right) \right] /$ $\left[ 4\kappa
\left( 1-\kappa \right)
\left( x_{0}^{2}+\tilde{\Delta}^{2}\right) ^{3/4}\right] $, $K\left( \tilde{%
\mu},\tilde{\Delta},\tilde{\lambda}\right) $
$=\tilde{\lambda}F\left( 2\theta ,\kappa \right) /$ $\left(
x_{0}^{2}+\tilde{\Delta}^{2}\right) ^{1/4}
$ where $x_{0}=\tilde{\mu}+\tilde{\lambda}^{2}$, $2\kappa =1+x_{0}/\sqrt{%
x_{0}^{2}+\tilde{\Delta}^{2}}$, $\tan \theta =\tilde{\lambda}/\left(
x_{0}^{2}+\tilde{\Delta}^{2}\right) ^{1/4}$, }$N\left( \phi ,\kappa
\right)
=E\left( \phi ,\kappa \right) +(\kappa -1)F\left( \phi ,\kappa \right) $, and%
{\ }$M\left( \phi ,\kappa \right) =\kappa \sin \phi \cos \phi /\sqrt{%
1-\kappa \sin ^{2}\phi }$. {$E\left( \phi ,\kappa \right) =\int_{0}^{\phi }%
\sqrt{1-\kappa \sin ^{2}x}dx$ and $F\left( \phi ,\kappa \right)
=\int_{0}^{\phi }1/\sqrt{1-\kappa \sin ^{2}x}dx$ are the first and
second kind incomplete elliptic integrals respectively with $0\leq
\phi \leq \pi $ in this paper.}
\end{thebibliography}
\end{document}